# Heliyon

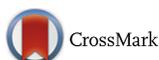

# Thermodynamic restrictions on linear reversible and irreversible thermo-electro-magneto-mechanical processes

**Sushma Santapuri** [a,b,*]

[a] Department of Applied Mechanics, Indian Institute of Technology Delhi, New Delhi, 110016, India

[b] Department of Mechanical Engineering, Polytechnic University of Puerto Rico, San Juan, 00918, Puerto Rico

* Correspondence to: Department of Applied Mechanics, Indian Institute of Technology Delhi, New Delhi, 110016, India.
E-mail address: ssantapuri@am.iitd.ac.in.

## Abstract

A unified thermodynamic framework for the characterization of functional materials is developed. This framework encompasses linear reversible and irreversible processes with thermal, electrical, magnetic, and/or mechanical effects coupled. The comprehensive framework combines the principles of classical equilibrium and non-equilibrium thermodynamics with electrodynamics of continua in the infinitesimal strain regime.

In the first part of this paper, linear Thermo-Electro-Magneto-Mechanical (TEMM) quasistatic processes are characterized. Thermodynamic stability conditions are further imposed on the linear constitutive model and restrictions on the corresponding material constants are derived. The framework is then extended to irreversible transport phenomena including thermoelectric, thermomagnetic and the state-of-the-art spintronic and spin caloritronic effects. Using Onsager's reciprocity relationships and the dissipation inequality, restrictions on the kinetic coefficients corresponding to charge, heat and spin transport processes are derived. All the constitutive models are accompanied by *multiphysics interaction diagrams* that highlight the various processes that can be characterized using this framework.

Keywords: Applied mathematics, Materials science, Thermodynamics

## 1. Introduction

Functional materials are engineered materials that are designed to exhibit desired functionalities (e.g., sensing, actuation, energy harvesting, self-healing) in response to a controllable stimulus. These materials have wide spread applications in fields like aerospace, automotive, medicine, electronics and defense [9, 39]. Some examples of such materials include multiferroic materials, bio-mimetic materials, semiconductors and spintronic materials.

Design and characterization of functional materials is at the forefront of materials research. These materials often exhibit coupling of various physical effects and are typically tailored to exhibit unusual electrical, magnetic, chemical, optical and/or thermal properties. In order to optimally design such materials the relationships between processing, structure, property, and performance of the material need to be established [15]. Such relationships are obtained through a combination of experiments, theory and computational models that range from atomic scale to macro/continuum scale [27, 28].

To this end, this paper aims to address one particular aspect of the characterization of functional materials, i.e., the development of an overarching thermodynamic framework that characterizes the thermal, electrical, magnetic and mechanical effects occurring in these materials. The early models presented in [13, 16, 18, 29, 38] for *fully coupled* Thermo-Electro-Magneto-Mechanical (TEMM) materials are used as a starting point in this work. These seminal mathematical models combine the principles of classical electrodynamics with thermomechanical conservation laws and can be applied to several materials ranging from linear piezoelectric materials, magnetostrictive materials to nonlinear electro-elastic solids, and electro-rheological fluids. Some of these applications were studied in [12, 17, 21, 22, 26, 30, 31, 37, 42].

In the more recent literature, unified thermodynamic models were developed to characterize a broader range of multiphysical processes. For instance, a thermo-electro-magnetic system with specific application to dielectric materials in the presence of memory effects was studied by Amendola [4] and the conditions for thermodynamic stability were investigated. This work was extended by Yu Li who studied the uniqueness and reciprocity of coupled thermo-electro-magneto-elastic behavior in smart materials [19]. More recently, Yu and Shen proposed a variational principle for coupled thermal-electrical-chemical-mechanical problems [41]. This framework described heat conduction, mass diffusion, electrochemical reactions and electrostatic processes. Characterization of dissipative functional materials with quasistatic electro-magneto-mechanical couplings was presented by Miehe et al. [23] based on incremental variational principles and stability analysis was performed on the macroscopic level based on the convexity/concavity of potentials. An internal variable based irreversible thermodynamics framework was formulated by Oates et al. [25] for ferroic materials which incorporated hysteretic behavior.

The examples discussed above demonstrate the theoretical development for a specific class of materials or processes. To this end, in this paper, a unified approach to thermodynamic modeling of a general thermo-electro-magneto-mechanical system is presented. Specifically, the first principles based thermodynamic framework developed in [32, 33] is utilized to characterize TEMM processes and subsequently specialized to model linear reversible and irreversible transport processes using classical equilibrium and non-equilibrium thermodynamics principles. Within the quasistatic regime, this work unifies all the known coupled and uncoupled TEMM processes and studies the stability conditions. Within the irreversible regime, this work unifies *memoryless* heat, charge, as well as the state-of-the-art spin transport phenomena to obtain a comprehensive set of modeling equations [7]. While this work does not deal with higher order effects like large deformation or hysteresis, these effects could be incorporated into the framework by adding additional independent variables and using similar characterization techniques.

This paper is structured as follows: Section 2 describes the first principle equations governing a fully coupled thermo-electro-magneto-mechanical medium in a small strain and small electromagnetic field regime. Section 3 deals with the development part of the thermodynamic framework wherein Section 3.1 describes the constitutive modeling of a near-equilibrium TEMM process. Conditions for stability of the thermodynamic equilibrium are investigated in 3.1.1. Utility of this framework is subsequently demonstrated through an example of a multiferroic material of hexagonal crystal symmetry wherein the restrictions on material constants are derived in Section 3.1.2. The framework is then extended to characterize irreversible transport processes in Section 3.2. These processes include the thermoelectric, galvanomagnetic/thermomagnetic and the spintronic/spin caloritronic effects. In Section 3.2.1, a modified version of the dissipation inequality is posited to incorporate the new variables corresponding to the spin transport phenomenon. The relationships between the various process constants are obtained through the use of *Onsager's reciprocity relationships* and the bounds on the process constants are derived from the dissipation inequality in Section 4. Finally, in Section 5 concluding remarks and the overall contributions of this work are discussed.

## 2. Background

## 2.1. Description of a fully coupled thermo-electro-magneto-mechanical process

In this section, the fundamental balance laws, governing the evolution of TEMM fields in a deformable, polarizable and magnetizable medium, are presented in the Cartesian component notation. These equations include the thermomechanical balance laws and the Maxwell's equations specialized to a small strain and small electromagnetic fields regime[1]:

$$\rho = \rho_R, \qquad \text{(Conservation of Mass)} \qquad (1a)$$

$$\rho u_{i,tt} = \rho f_i^m + T_{ij,j}, \qquad \text{(Balance of Linear Momentum)} \qquad (1b)$$

$$T_{ij} = T_{ji}, \qquad \text{(Balance of Angular Momentum)} \qquad (1c)$$

$$\rho \varepsilon_{,t} = T_{ij} E_{ij,t} + e_i p_{i,t} + \mu_o h_i m_{i,t}$$
$$+ \rho r^t + j_i e_i - q_{i,i}, \qquad \text{(First Law of Thermodynamics)} \qquad (1d)$$

$$\rho \eta_{,t} \geq \rho \frac{r^t}{\Theta} - \left(\frac{q_i}{\Theta}\right)_{,i}, \qquad \text{(Second Law of Thermodynamics)} \qquad (1e)$$

$$b_{i,i} = 0, \qquad \text{(Gauss's Law for Magnetism)} \qquad (1f)$$

$$b_{i,t} + \varepsilon_{ijk} e_{k,j} = 0, \qquad \text{(Faraday's Law)} \qquad (1g)$$

$$d_{i,i} = \sigma, \qquad \text{(Gauss's Law for Electricity)} \qquad (1h)$$

$$d_{i,t} + j_i = \varepsilon_{ijk} h_{k,j}. \qquad \text{(Ampere-Maxwell Law)} \qquad (1i)$$

The notation $(\ )_{,t}$ denotes partial differentiation with respect to time, e.g.,

$$b_{i,t} \equiv \frac{\partial b_i(x_1, x_2, x_3, t)}{\partial t}.$$

The TEMM fields appearing in (1a)–(1i) include the density $\rho$, specific internal energy $\varepsilon$ (internal energy per unit mass), specific entropy $\eta$, the absolute temperature $\Theta$, the thermally and electromagnetically induced specific energy supply rates $r^t$ and $r^{em}$ and the Cartesian components of the displacement $u_i$, the Cauchy stress tensor $T_{ij}$, the specific body force $f_i^m$ and the heat flux vector $q_i$. Also, $e_i, d_i, h_i, b_i, \sigma$, and $j_i$ represent the Cartesian components of electric field intensity, electric displacement, magnetic field intensity, magnetic induction, free charge density, and free current density, respectively. Additionally,

$$p_i = d_i - \epsilon_o e_i, \qquad m_i = \frac{1}{\mu_o} b_i - h_i, \qquad (2)$$

are the Cartesian components of the electric polarization and magnetization vectors. Also, $\epsilon_o$ and $\mu_o$ are the permittivity and permeability constants *in vacuo*. Finally, the infinitesimal strain tensor $E_{ij}$ is related to the displacement $u_i$ as

$$E_{ij} = \frac{1}{2}\left(u_{i,j} + u_{j,i}\right). \tag{3}$$

---

[1] Derivation of the small strain theory of TEMM materials is presented in [32] (cf. Section 9.7.1).

The Cauchy stress (and strain) tensors are typically non-symmetric in the presence of electromagnetically induced body force and body couple. However, their contributions to the balance of linear momentum (1b) and angular momentum (1c) emerge at higher orders and can be ignored for small electromagnetic fields.

To complete the mathematical model, the governing equations need to be supplemented with the material specific constitutive equations as well as the boundary conditions. In the subsequent sections, constitutive equations will be developed for fully coupled TEMM reversible and irreversible processes operating in the small strain, small EM (electromagnetic) field regime. Furthermore, the ramifications of the second law of thermodynamics as well as the thermodynamic stability restrictions on the proposed constitutive models will be studied.

## 2.2. Characterization of quasistatic thermo-electro-magneto-mechanical material processes

The development of a continuum thermodynamic framework for fully coupled TEMM materials was presented in [33]. The principles of classical thermodynamics and electrodynamics of continua were utilized to develop the thermodynamics state equations corresponding to various combinations of independent variables. In this paper, starting from the state equations derived in [33], constitutive models are developed for TEMM materials in a linear regime.

As a starting point, the reduced form of the Clausius–Duhem inequality (obtained by combining the first law of thermodynamics (1d) and the second law of thermodynamics (1e) via the elimination of $r^t$) is presented below:

$$-\rho\,\dot{\varepsilon} + \underbrace{\mathbf{T}\cdot\dot{\mathbf{E}} + \rho\,\theta\,\dot{\eta} + \mathbf{e}\cdot\dot{\mathbf{p}} + \mu_o\,\mathbf{h}\cdot\dot{\mathbf{m}}}_{\text{TEMM Conjugate Variables}} + \mathbf{j}\cdot\mathbf{e} - \frac{1}{\theta}\,\mathbf{q}\cdot\operatorname{grad}\theta \;\geq\; 0. \tag{4}$$

The reduced Clausius–Duhem inequality can be utilized to develop the thermodynamic state equations corresponding to any free energy with a desired combination of independent variables, as demonstrated in [33]. In this paper, the internal energy based formulation, namely, the free energy

$$E^{F\eta pm} \equiv \varepsilon(\mathbf{E}, \eta, \mathbf{p}, \mathbf{m}) \tag{5}$$

characterized by infinitesimal strain **E**, entropy $\eta$, polarization **p** and magnetization **m** as independent variables will be utilized to develop the constitutive formulation.[2] The corresponding thermodynamic state equations as demonstrated in [33] include

$$\mathbf{T} \;=\; \rho \,\frac{\partial \varepsilon}{\partial \mathbf{E}}, \qquad \theta \;=\; \frac{\partial \varepsilon}{\partial \eta}, \qquad \mathbf{e} \;=\; \rho \,\frac{\partial \varepsilon}{\partial \mathbf{p}}, \qquad \mathbf{h} \;=\; \frac{\rho}{\mu_o} \,\frac{\partial \varepsilon}{\partial \mathbf{m}}. \tag{6}$$

## 3. Methodology

### 3.1. Constitutive model development I: quasistatic TEMM processes

In what follows, starting from the state equations (6), linear TEMM constitutive equations are formulated in the near-equilibrium regime. The functional form of internal energy $\varepsilon = E^{F\eta pm}$ is derived by performing a Taylor series expansion of the free energy in terms of its independent variables in the neighborhood of a thermodynamic equilibrium state $\mathbf{x}_o = (\mathbf{E}^o, \eta^o, \mathbf{p}^o, \mathbf{m}^o)$:

$$
\begin{aligned}
\breve{E}^{F\eta pm} \;=\;& \varepsilon\,(\mathbf{E}, \eta, \mathbf{p}, \mathbf{m}) \\[4pt]
=\;& \varepsilon|_{x_o} + \frac{1}{2}\,\frac{\partial^2 \varepsilon}{\partial E_{ij}\partial E_{kl}}\bigg|_{x_o} E_{ij}E_{kl} + \frac{1}{2}\,\frac{\partial^2 \varepsilon}{\partial p_i \partial p_j}\bigg|_{x_o} p_i p_j \\[6pt]
&+ \frac{1}{2}\,\frac{\partial^2 \varepsilon}{\partial m_i \partial m_j}\bigg|_{x_o} m_i m_j + \frac{1}{2}\,\frac{\partial^2 \varepsilon}{\partial \eta^2}\bigg|_{x_o} \eta^2 + \frac{\partial^2 \varepsilon}{\partial E_{ij}\partial p_k}\bigg|_{x_o} E_{ij}p_k \\[6pt]
&+ \frac{\partial^2 \varepsilon}{\partial E_{ij}\partial m_k}\bigg|_{x_o} E_{ij}m_k + \frac{\partial^2 \varepsilon}{\partial p_i \partial m_j}\bigg|_{x_o} p_i m_j + \frac{\partial^2 \varepsilon}{\partial \eta \partial E_{ij}}\bigg|_{x_o} \eta\, E_{ij} \\[6pt]
&+ \frac{\partial^2 \varepsilon}{\partial \eta \partial p_i}\bigg|_{x_o} \eta\, p_i + \frac{\partial^2 \varepsilon}{\partial \eta \partial m_i}\bigg|_{x_o} \eta\, m_i + \text{higher order terms}, \tag{7}
\end{aligned}
$$

where $E_{ij}$, $p_i$ and $m_i$ are the Cartesian components of the infinitesimal strain tensor **E**, polarization **p**, and magnetization **m**, and $\eta$ is the entropy of the system perturbed about the initial state $\mathbf{x}_o$. The linear TEMM constitutive equations are derived by substituting the free energy function (7) into the state equations (6) and ignoring the higher order terms:

$$T_{ij} \;=\; \frac{\partial^2 \bar{\varepsilon}}{\partial E_{ij}\partial E_{kl}}\bigg|_{x_o} E_{kl} + \frac{\partial^2 \bar{\varepsilon}}{\partial E_{ij}\,\partial p_k}\bigg|_{x_o} p_k + \frac{\partial^2 \bar{\varepsilon}}{\partial E_{ij}\partial m_k}\bigg|_{x_o} m_k + \frac{\partial^2 \bar{\varepsilon}}{\partial \bar{\eta}\,\partial E_{ij}}\bigg|_{x_o} \bar{\eta}, \tag{8}$$

$$e_i \;=\; \frac{\partial^2 \bar{\varepsilon}}{\partial E_{jk}\partial p_i}\bigg|_{x_o} E_{jk} + \frac{\partial^2 \bar{\varepsilon}}{\partial p_k \partial p_i}\bigg|_{x_o} p_k + \frac{\partial^2 \bar{\varepsilon}}{\partial p_i \partial m_k}\bigg|_{x_o} m_k + \frac{\partial^2 \bar{\varepsilon}}{\partial \bar{\eta}\,\partial p_i}\bigg|_{x_o} \bar{\eta}, \tag{9}$$

---

[2] Equivalent thermodynamic models could be developed for other free energy formulations using *Legendre transforms* and utilizing the Coleman and Noll approach [11].

**Table 1.** Materials constants and their representations for linear reversible processes.

| Constant | Representation | Constant | Representation |
|---|---|---|---|
| Elasticity Constant | $C_{ijkl} = \dfrac{\partial^2 \varepsilon}{\partial E_{ij} \partial E_{kl}}\Big|_{x_o}$ | Piezoelectric Constant | $g^e_{ijk} = \dfrac{\partial^2 \varepsilon}{\partial E_{ij} \partial p_k}\Big|_{x_o}$ |
| Piezomagnetic Constant | $g^m_{ijk} = \dfrac{\partial^2 \varepsilon}{\partial E_{ij} \partial m_k}\Big|_{x_o}$ | Coefficient of Thermal Stress | $\beta_{ij} = \dfrac{\partial^2 \varepsilon}{\partial \eta \partial E_{ij}}\Big|_{x_o}$ |
| Inverse Electric Susceptibility | $K^e_{ij} = \dfrac{\partial^2 \varepsilon}{\partial p_i \partial p_j}\Big|_{x_o}$ | Magneto-Electric Constant | $K^{em}_{ij} = \dfrac{\partial^2 \varepsilon}{\partial p_i \partial m_j}\Big|_{x_o}$ |
| Pyroelectric Constant | $L^e_i = \dfrac{\partial^2 \varepsilon}{\partial p_i \partial \eta}\Big|_{x_o}$ | Inverse Magnetic Susceptibility | $K^m_{ij} = \dfrac{\partial^2 \varepsilon}{\partial m_i \partial m_j}\Big|_{x_o}$ |
| Pyromagnetic Constant | $L^m_i = \dfrac{\partial^2 \varepsilon}{\partial m_i \partial \eta}\Big|_{x_o}$ | Inverse Specific Heat | $c = \dfrac{\partial^2 \varepsilon}{\partial \eta^2}\Big|_{x_o}$ |

$$h_i = \frac{\partial^2 \bar{\varepsilon}}{\partial E_{jk} \partial m_i}\Big|_{x_o} E_{jk} + \frac{\partial^2 \bar{\varepsilon}}{\partial m_i \partial p_k}\Big|_{x_o} p_k + \frac{\partial^2 \bar{\varepsilon}}{\partial m_k \partial m_i}\Big|_{x_o} m_k + \frac{\partial^2 \bar{\varepsilon}}{\partial \bar{\eta} \partial m_i}\Big|_{x_o} \bar{\eta}, \quad (10)$$

$$\theta = \frac{\partial^2 \bar{\varepsilon}}{\partial E_{ij} \partial \bar{\eta}}\Big|_{x_o} E_{ij} + \frac{\partial^2 \bar{\varepsilon}}{\partial p_i \partial \bar{\eta}}\Big|_{x_o} p_i + \frac{\partial^2 \bar{\varepsilon}}{\partial m_i \partial \bar{\eta}}\Big|_{x_o} m_i + \frac{\partial^2 \bar{\varepsilon}}{\partial \bar{\eta}^2}\Big|_{x_o} \bar{\eta}, \quad (11)$$

where $\bar{\varepsilon} = \rho\,\varepsilon$ and $\bar{\eta} = \rho\,\eta$ represent the internal energy per unit volume and the entropy per unit volume of the system, respectively. The coefficients arising in the linear constitutive equations (8)–(11) are material specific constants corresponding to different TEMM processes (described in Figure 1). For instance, the coefficient $\left(\partial^2 \bar{\varepsilon}/\partial E_{ij} \partial E_{kl}\right)\Big|_{x_o}$ can be identified as the stiffness matrix or elasticity constant of a material. In Table 1, the nomenclature of all the coefficients is presented. Throughout the subsequent development, internal energy per unit volume and entropy per volume are used and the bars are dropped for a simplified presentation.

The resulting free energy function for a fully coupled linear TEMM process is

$$E^{F\eta pm} = \frac{1}{2} C_{ijkl}\, E_{ij} E_{kl} + \frac{1}{2} K^e_{ij}\, p_i p_j + \frac{1}{2} K^m_{ij}\, m_i m_j + \frac{1}{2}\, c\,\eta^2 + g^e_{ijk}\, E_{ij} p_k$$
$$+ g^m_{ijk}\, E_{ij} m_k + \beta_{ij}\, E_{ij}\eta + K^{em}_{ij}\, p_i m_j + L^e_i\, \eta p_i + L^m_i\, \eta m_i. \quad (12)$$

The class of materials and coupled processes that can be characterized using this linear framework are described through the *Multiphysics Interaction Diagram* (MPID) shown in Figure 1. This diagram identifies all the known reversible thermo-electro-magneto-mechanical processes [20, 32]. Specifically, the TEMM extensive variables are marked on the corners of the inner quadrilateral and the intensive variables are marked on the outside. The green lines highlight the TEMM processes that couple any two of the four physical effects, whereas the blue lines represent the uncoupled processes. Furthermore, the arrows denote the direction of the processes. For example, *piezoelectricity*, defined as the accumulation of electric charge in response to an applied stress, is represented in the MPID by the green line that connects the electric polarization **p** and the Cauchy stress **T**. The direction of the arrow signifies the generation of material polarization (effect) in response to an applied mechanical stress (cause).

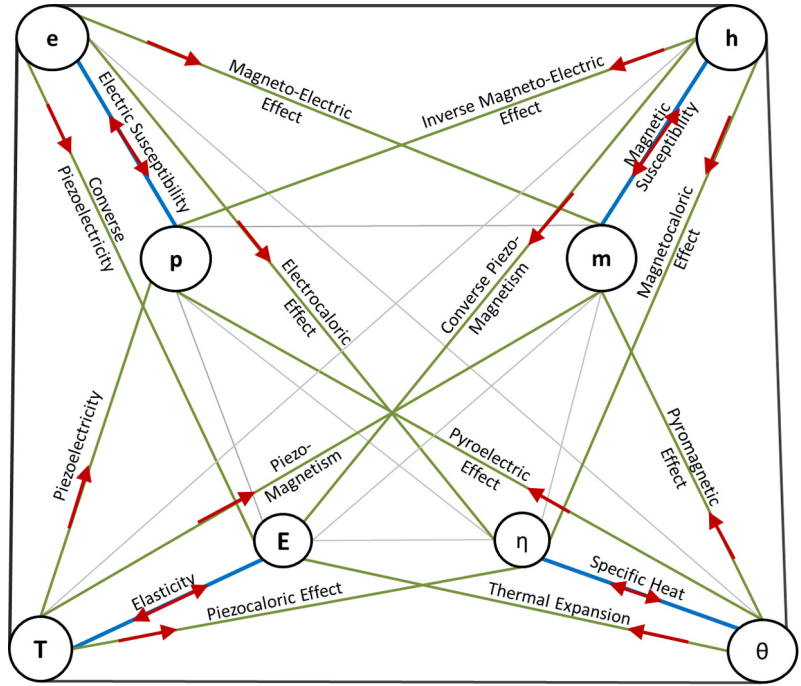

**Figure 1.** Multiphysics interaction diagram demonstrating linear thermo-electro-magneto-mechanical effects [32].

As stated earlier, within the infinitesimal strain regime, the stress and strain tensors are symmetric, i.e., they have only 6 independent components. This allows us to simplify the representation of the stress and strain tensors as well as the corresponding material constants using the Voigt notation, wherein the tensor indices are replaced as shown below:

$$11 \rightarrow 1, \quad 22 \rightarrow 2, \quad 33 \rightarrow 3, \quad 12,21 \rightarrow 4, \quad 23,32 \rightarrow 5, \quad 13,31 \rightarrow 6.$$

Using this shorthand notation, the constitutive equations (8)–(11) can be simplified further

$$T_I = C_{IJ} E_J + g_{Ij}^e p_j + g_{Ij}^m m_j + \beta_I \eta, \tag{13a}$$

$$e_i = g_{Ji}^e E_J + K_{ij}^e p_j + K_{ij}^{em} m_j + L_i^e \eta, \tag{13b}$$

$$\mu_o h_i = d_{Ji}^m E_J + K_{ij}^{em} p_j + K_{ij}^m m_j + L_i^m \eta, \tag{13c}$$

$$\theta = \beta_I E_I + L_i^e p_i + L_i^m m_i + c \eta, \tag{13d}$$

where $I, J \in 1, 2..., 6$ and $i, j \in 1, 2, 3$.

The linear model presented here characterizes a class of materials called ferroic materials that exhibit spontaneous polarization or magnetization in the presence of external electromagnetic fields. It is noted that the constitutive models developed here have a limited regime of applicability, i.e., within a small perturbation of an equilibrium state, often approximated as a linear, reversible process. Thus, the linearized constitutive models will not be able to predict effects like nonlinearity, irreversibility, dissipation or large deformations. For instance, *piezomagnetism*, i.e., the magneto-mechanical coupling effect occurring in ferromagnetic materials, is a linear approximation of *magnetostriction* which is in fact a highly nonlinear and hysteretic (irreversible) effect. In order to predict the complete nonlinear regime of magnetostriction accurately, additional independent variables, known as the *internal variables*, need to be used to describe the microstructural evolution and the associated losses in a material at lower scales.

### 3.1.1. Thermodynamic stability

Any spontaneous change in the parameters of a system in stable equilibrium will result in processes that aim to restore the system to its prior equilibrium state [36]. In other words, a thermodynamically stable system cannot grow rapidly from small perturbations about the equilibrium.

In this section, we look into the conditions required for such a stable equilibrium state. A consequence of this requirement is that the internal energy of the material must be a convex function of the extensive variables, which is imposed as follows [6]:

Convexity of $\varepsilon(\mathbf{E}, \eta, \mathbf{p}, \mathbf{m})$

$$\Leftrightarrow \ \mathbf{H} = \begin{bmatrix} \frac{\partial^2 \varepsilon}{\partial E_I \partial E_J} & \frac{\partial^2 \varepsilon}{\partial E_I \partial p_j} & \frac{\partial^2 \varepsilon}{\partial E_I \partial m_j} & \frac{\partial^2 \varepsilon}{\partial E_I \partial \eta} \\ \frac{\partial^2 \varepsilon}{\partial p_i \partial E_J} & \frac{\partial^2 \varepsilon}{\partial p_i \partial p_j} & \frac{\partial^2 \varepsilon}{\partial p_i \partial m_j} & \frac{\partial^2 \varepsilon}{\partial p_i \partial \eta} \\ \frac{\partial^2 \varepsilon}{\partial m_i \partial E_J} & \frac{\partial^2 \varepsilon}{\partial m_i \partial p_j} & \frac{\partial^2 \varepsilon}{\partial m_i \partial m_j} & \frac{\partial^2 \varepsilon}{\partial m_i \partial \eta} \\ \frac{\partial^2 \varepsilon}{\partial \eta \partial E_J} & \frac{\partial^2 \varepsilon}{\partial \eta \partial p_j} & \frac{\partial^2 \varepsilon}{\partial \eta \partial m_j} & \frac{\partial^2 \varepsilon}{\partial \eta^2} \end{bmatrix} \text{ is positive definite}$$

for all possible values of independent variables $\mathbf{E}$, $\eta$, $\mathbf{p}$, and $\mathbf{m}$ within a small perturbation of the equilibrium state. $\mathbf{H}$ is the *Hessian matrix* corresponding to the linear constitutive equations (8)–(11) and can be expressed as a *block matrix* consisting of all the coefficient matrices defined in Table 1:

$$\mathbf{H} = \begin{bmatrix} \mathbf{C} & \mathbf{g}^e & \mathbf{g}^m & \boldsymbol{\beta} \\ (\mathbf{g}^e)^T & \mathbf{K}^e & \mathbf{K}^{em} & \mathbf{L}^e \\ (\mathbf{g}^m)^T & (\mathbf{K}^{em})^T & \mathbf{K}^m & \mathbf{L}^m \\ (\boldsymbol{\beta})^T & (\mathbf{L}^e)^T & (\mathbf{L}^m)^T & c \end{bmatrix}. \tag{14}$$

A necessary (but not sufficient) condition for convexity of $\mathbf{H}$ is the Legendre–Hadamard condition

$$det(\mathbf{H}) \geq 0. \tag{15}$$

In the following section, we demonstrate the application of these restrictions on a linear *multiferroic material* with a specified crystallographic symmetry.

### 3.1.2. Example: multiferroic material with hexagonal symmetry

A multiferroic material exhibits coupling of two or more ferroic orders. In this example, a general multiferroic material with hexagonal crystal symmetry that exhibits a fully coupled TEMM behavior is considered. The linear TEMM constitutive equations (8)–(11) reduce to the following form for hexagonal symmetry (6mm crystallographic symmetry and 6m'm' magnetic point symmetry)[3]:

$$
\begin{bmatrix} T_1 \\ T_2 \\ T_3 \\ T_4 \\ T_5 \\ T_6 \\ e_1 \\ e_2 \\ e_3 \\ h_1 \\ h_2 \\ h_3 \\ \theta \end{bmatrix} =
\underbrace{\begin{bmatrix}
C_{11} & C_{12} & C_{13} & 0 & 0 & 0 & 0 & 0 & g_{13}^e & 0 & 0 & g_{13}^m & \beta_1 \\
C_{12} & C_{11} & C_{13} & 0 & 0 & 0 & 0 & 0 & g_{13}^e & 0 & 0 & g_{13}^m & \beta_1 \\
C_{13} & C_{13} & C_{33} & 0 & 0 & 0 & 0 & 0 & g_{33}^e & 0 & 0 & g_{33}^m & \beta_3 \\
0 & 0 & 0 & C_{44} & 0 & 0 & g_{41}^e & g_{51}^e & 0 & 0 & g_{51}^m & 0 & 0 \\
0 & 0 & 0 & 0 & C_{44} & 0 & g_{51}^e & g_{41}^e & 0 & g_{51}^m & 0 & 0 & 0 \\
0 & 0 & 0 & 0 & 0 & C_{66} & 0 & 0 & 0 & 0 & 0 & 0 & 0 \\
0 & 0 & 0 & g_{41}^e & g_{51}^e & 0 & K_{11}^e & 0 & 0 & K_{11}^{em} & 0 & 0 & 0 \\
0 & 0 & 0 & g_{51}^e & g_{41}^e & 0 & 0 & K_{11}^e & 0 & 0 & K_{11}^{em} & 0 & 0 \\
g_{13}^e & g_{13}^e & g_{33}^e & 0 & 0 & 0 & 0 & 0 & K_{33}^e & 0 & 0 & K_{33}^{em} & L_3^e \\
0 & 0 & 0 & g_{51}^m & 0 & 0 & K_{11}^{em} & 0 & 0 & K_{11}^m & 0 & 0 & 0 \\
0 & 0 & 0 & g_{51}^m & 0 & 0 & 0 & K_{11}^{em} & 0 & 0 & K_{11}^m & 0 & 0 \\
g_{13}^m & g_{13}^m & g_{33}^m & 0 & 0 & 0 & 0 & 0 & K_{33}^{em} & 0 & 0 & K_{33}^m & L_3^m \\
\beta_1 & \beta_1 & \beta_3 & 0 & 0 & 0 & 0 & 0 & L_3^e & 0 & 0 & L_3^m & c
\end{bmatrix}}_{\text{Hessian Matrix } \mathbf{H}}
\begin{bmatrix} E_1 \\ E_2 \\ E_3 \\ E_4 \\ E_5 \\ E_6 \\ p_1 \\ p_2 \\ p_3 \\ \mu_o m_1 \\ \mu_o m_2 \\ \mu_o m_3 \\ \eta \end{bmatrix}
\tag{16}
$$

where $C_{66} = 1/2(C_{11} - C_{12})$. All the material constants follow the same notation as defined in Table 1.

For the linear constitutive model (16), the coefficient matrix coincides with the Hessian **H**. Restricting the determinant of **H** to positive values generates bounds on the material constants. This condition must be valid for any combination of the TEMM independent variables. Thus, the stability requirements corresponding to some of the special cases are presented below:

- For $\mathbf{E} \neq \mathbf{0}$, $\mathbf{p} = \mathbf{m} = \mathbf{0}$, $\eta = 0$, stability restrictions reduce to

$$
C_{11} \geq 0, \quad C_{44} \geq 0, \quad -C_{11} \leq C_{12} \leq C_{11}, \quad 2(C_{13})^2 \leq C_{22}(C_{11} + C_{12}).
\tag{17}
$$

- For $\mathbf{E} \neq \mathbf{0}$, $\mathbf{p} \neq \mathbf{0}$, $\mathbf{m} = \mathbf{0}$, $\eta = 0$, we obtain

$$
K_{11}^e \geq 0, \quad K_{33}^e \geq 0, \quad (g_{51}^e)^2 \leq K_{11}^e C_{44}, \quad 2(g_{13}^e)^2 \leq K_{33}^e(C_{11} + C_{12}).
\tag{18}
$$

---

[3] The coefficient matrices corresponding to hexagonal symmetry are derived in [24].

- Similarly, for $\mathbf{E} \neq \mathbf{0}$, $\mathbf{m} \neq \mathbf{0}$, $\mathbf{p} = \mathbf{0}$, $\eta = 0$, we obtain

$$K_{11}^m \geq 0, \quad K_{33}^m \geq 0, \quad (g_{51}^m)^2 \leq K_{11}^m C_{44}, \quad 2(g_{13}^m)^2 \leq K_{33}^m (C_{11} + C_{12}).$$
(19)

- Now considering $\mathbf{p} \neq \mathbf{0}$, $\mathbf{m} \neq \mathbf{0}$, $\mathbf{E} = \mathbf{0}$, $\eta = 0$, we get

$$(K_{11}^{em})^2 \leq K_{11}^e K_{11}^m, \quad (K_{33}^{em})^2 \leq K_{33}^e K_{33}^m.$$
(20)

- Also, for $\eta \neq 0$, $\mathbf{E} \neq \mathbf{0}$, $\mathbf{p} = \mathbf{m} = \mathbf{0}$, we have

$$c \geq 0, \quad (\beta_3)^2 \leq c\, C_{33}, \quad (\beta_1)^2 \leq c\, C_{11}.$$
(21)

- Finally, considering $\eta \neq 0$, $\mathbf{p} \neq \mathbf{0}$ or $\mathbf{m} \neq \mathbf{0}$, we obtain

$$(L_3^e)^2 \leq c\, K_{33}^e, \quad (L_3^m)^2 \leq c\, K_{33}^m.$$
(22)

In what follows, we extend the framework to characterize irreversible processes characterized by heat, charge and spin transport.

## 3.2. Constitutive model development II: transport processes

The framework presented in Section 3.1 assumes a slow and thermodynamically reversible process. In this section, characterization of irreversible transport processes (associated with rates and gradients of physical quantities) will be developed starting from the Clausius–Duhem inequality (4) and utilizing irreversible thermodynamics principles.

### 3.2.1. Characterization of transport processes

Revisiting the reduced Clausius–Duhem inequality (4):

$$-\dot{\varepsilon} + \mathbf{T} \cdot \dot{\mathbf{E}} + \theta\dot{\eta} + \mathbf{e} \cdot \dot{\mathbf{p}} + \mu_o\, \mathbf{h} \cdot \dot{\mathbf{m}} + \mathbf{j} \cdot \mathbf{e} - \frac{1}{\theta}\mathbf{q} \cdot \mathrm{grad}\,\theta \geq 0$$
(23)

where $\varepsilon$ and $\eta$ correspond to the internal energy and entropy per unit volume. While the quasistatic processes are characterized in terms of the TEMM extensive–intensive conjugate variables, the transport processes are characterized in terms of the *thermodynamic forces* that drive the process, and the resulting *thermodynamic flow* terms that are generated as a response to the input forces. The free energy formulation described in Section 3.1 is thus extended to include additional independent and dependent variables, i.e.,

$$\varepsilon = \tilde{\varepsilon}(\mathbf{F}, \mathbf{p}, \mathbf{m}, \eta, \mathrm{grad}\,\theta, \mathrm{grad}\,\mu), \quad \mathbf{q} = \tilde{\mathbf{q}}(\mathbf{F}, \mathbf{p}, \mathbf{m}, \eta, \mathrm{grad}\,\theta, \mathrm{grad}\,\mu),$$

$$\mathbf{j} = \tilde{\mathbf{j}}(\mathbf{F}, \mathbf{p}, \mathbf{m}, \eta, \mathrm{grad}\,\theta, \mathrm{grad}\,\mu),$$
(24)

wherein the gradient of temperature $\text{grad}\,\theta$ and the gradient of electrochemical potential $\text{grad}\,\mu$ are added as the independent variables (i.e., the *thermodynamic forces*) whereas the electric current density $\mathbf{j}$ and the heat current density $\mathbf{q}$ are added as the dependent variables (i.e., the *thermodynamic flow terms*). As is customary, we now apply the chain rule on the free energy function (24)$_1$:

$$\dot{\varepsilon} = \frac{\partial \varepsilon}{\partial \mathbf{F}} \cdot \dot{\mathbf{F}} + \frac{\partial \varepsilon}{\partial \mathbf{p}} \cdot \dot{\mathbf{p}} + \frac{\partial \varepsilon}{\partial \mathbf{m}} \cdot \dot{\mathbf{m}} + \frac{\partial \varepsilon}{\partial \eta} \cdot \dot{\eta} + \frac{\partial \varepsilon}{\partial (\text{grad}\,\theta)} \cdot \overline{\text{grad}\,\theta} + \frac{\partial \varepsilon}{\partial (\text{grad}\,\mu)} \cdot \overline{\text{grad}\,\mu}.$$
(25)

Substituting in the Clausius–Duhem inequality and using the Coleman and Noll approach [10], we obtain

$$\left( \mathbf{P} - \frac{\partial \varepsilon}{\partial \mathbf{F}} \right) \cdot \dot{\mathbf{F}} + \left( \theta - \frac{\partial \varepsilon}{\partial \eta} \right) \cdot \dot{\eta} + \left( \mathbf{e} - \frac{\partial \varepsilon}{\partial \mathbf{p}} \right) \cdot \dot{\mathbf{p}} + \left( \mu_o \, \mathbf{h} - \frac{\partial \varepsilon}{\partial \mathbf{m}} \right) \cdot \dot{\mathbf{m}}$$

$$+ \frac{\partial \varepsilon}{\partial (\text{grad}\,\theta)} \cdot \overline{\text{grad}\,\theta} + \frac{\partial \varepsilon}{\partial (\text{grad}\,\mu)} \cdot \overline{\text{grad}\,\mu} + \frac{1}{\theta} \mathbf{j} \cdot \mathbf{e} - \mathbf{q} \cdot \frac{\text{grad}\,\theta}{\theta^2} \geq 0.$$
(26)

Since the rates $\dot{\mathbf{F}}, \dot{\eta}, \dot{\mathbf{p}}$, and $\dot{\mathbf{m}}$ are mutually independent and may be varied arbitrarily, it follows from (26) that the coefficients of the rates must vanish, i.e.,

$$\mathbf{P} = \frac{\partial \varepsilon}{\partial \mathbf{F}}, \qquad \theta = \frac{\partial \varepsilon}{\partial \eta}, \qquad \mathbf{e} = \frac{\partial \varepsilon}{\partial \mathbf{p}}, \qquad \mathbf{h} = \frac{1}{\mu_o} \frac{\partial \varepsilon}{\partial \mathbf{m}},$$
(27a)

$$\frac{\partial \varepsilon}{\partial (\text{grad}\,\theta)} = \mathbf{0}, \qquad \frac{\partial \varepsilon}{\partial (\text{grad}\,\mu)} = \mathbf{0},$$
(27b)

along with the residual inequality,

$$\mathbf{j} \cdot \mathbf{e} - \mathbf{q} \cdot \frac{\text{grad}\,\theta}{\theta} \geq 0.$$
(28)

It is evident from (27b) that free energy $\varepsilon$ is independent of $\text{grad}\,\theta$ and $\text{grad}\,\mu$. Thus, the irreversible transport processes are characterized using the functional descriptions for $\mathbf{j}$ and $\mathbf{q}$ of the form (24)$_{2,3}$ and subsequently restricted by the residual inequality (28).

### 3.2.2. Classification of transport processes

Three types of *memoryless* transport processes are studied in this paper, namely,

1. **Thermoelectric processes**: The transport phenomena associated with the flow of electric current and heat current in the *absence* of external magnetic field. These include physical effects like *heat conductivity, electrical conductivity, Seebeck effect* and *Peltier effect*.

2. **Thermomagnetic and Galvanomagnetic processes**: The transport processes that arise in the presence of an externally applied magnetic field. These effects are a result of the Lorentz forces acting on the moving free electrons, which in turn are generated due to the thermal or electrical potential gradients orthogonal to the applied magnetic field. Examples include *Nernst effect, Ettinghausen effect, Hall effect* and *Righi–Leduc effect (or Thermal Hall effect)*.

3. **Spin-induced processes or Spintronics**: The transport processes resulting from a net polarization of the spin-up and the spin-down electrons.[4] In most materials, electron spins are equally present in both the up ($s = +1/2$) and the down ($s = -1/2$) states. An imbalance between these states can be created by putting a magnetic material in a large magnetic field (Zeeman effect) or by utilizing the exchange energy present in a ferromagnet [14, 35]. Motion of such a spin-polarized population of electrons can result in a plethora of spin-induced transport phenomena that include *the Spin Hall effect, the Inverse Spin Hall effect, Spin-dependent Seebeck effect, Spin-dependent Peltier effect* and *the Spin Nernst effect*.

The multiphysics interaction diagrams (MPID) corresponding to all the transport processes described above are highlighted in the irreversible multiphysics interaction diagram, Figure 2. The irreversible MPID is divided into two parts wherein Figure 2(a) describes the purely thermoelectric transport processes (i.e., no magnetic field or spin polarization), whereas Figure 2(b) describes the thermomagnetic, galvanomagnetic and spintronic transport processes (i.e., non-zero external magnetic field or spin-polarization). The flow terms are marked on the inside and the thermodynamic force terms are marked on the outside. Similar to Figure 1, the individual processes are represented by lines connecting the appropriate flow-force quantities and the arrow signifies the direction of the process. Also, the uncoupled processes are represented by the blue lines whereas the coupled processes are marked in green.

In what follows, the characterization of transport processes will be modified to incorporate additional variables arising from spin-polarization of the electron population.

### 3.2.3. Characterization of spin transport

The thermodynamic formulation presented in Section 3.2.1 does not characterize spin transport. In order to extend the formulation to spin-dependent processes, additional independent and dependent variables are required for the complete

---

[4] Spin of an electron is associated with its intrinsic angular momentum, which is different from the angular momentum generated by the electron orbital motion. Experimental evidence suggests that the spin of electron can exist in two possible states, namely, the spin-up and the spin-down states.

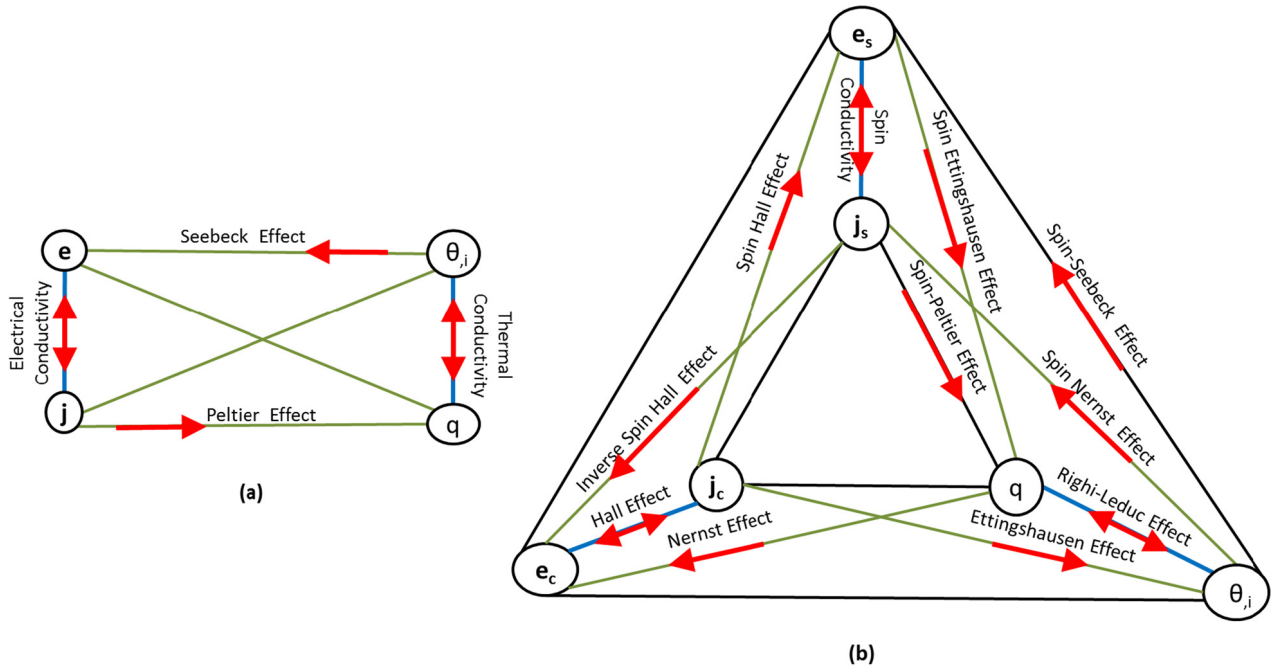

**Figure 2.** Multiphysics interaction diagram demonstrating (a) thermoelectric transport processes in the absence of external magnetic fields, and (b) galvanomagnetic, thermomagnetic, and spin-induced transport processes in the presence of external magnetic fields.

characterization. The following spin-dependent current and force quantities are thus defined:

- The **charge and spin-induced currents** defined by

$$\mathbf{j}_c = \mathbf{j}_\uparrow + \mathbf{j}_\downarrow \text{ and } \mathbf{j}_s = \mathbf{j}_\uparrow - \mathbf{j}_\downarrow, \tag{29}$$

where $\mathbf{j}_\uparrow$ and $\mathbf{j}_\downarrow$ denote the currents generated due to the motion of the spin-up and the spin-down charges, respectively. Also, $\mathbf{j}_c$ and $\mathbf{j}_s$ are defined as the charge current and the spin-polarized current, respectively [5]. In a spin-independent system $\mathbf{j}_\uparrow = \mathbf{j}_\downarrow$, which leads to zero spin-currents.

- The **charge and spin chemical potentials** are defined as [5]

$$\mu_c = \frac{\mu_\uparrow + \mu_\downarrow}{2}, \text{ and } \mu_s = \mu_\uparrow - \mu_\downarrow \tag{30}$$

such that

$$\frac{\text{grad } \mu_c}{e} = \mathbf{e}_c, \qquad \frac{\text{grad } \mu_s}{2e} = \mathbf{e}_s, \tag{31}$$

wherein $\mu_\uparrow$ and $\mu_\downarrow$ denote the electrochemical potentials induced by the motion of the spin-up and the spin-down electrons, respectively. Also, $\mathbf{e}_s$ and $\mathbf{e}_c$ are defined as the spin-induced and charge-induced electric fields, respectively.

### 3.2.4. Modified dissipation inequality

The modified form of dissipation inequality (28) that incorporates spin transport is now posited as

$$\mathbf{j}_\uparrow \cdot \mathbf{e}_\uparrow + \mathbf{j}_\downarrow \cdot \mathbf{e}_\downarrow - \frac{1}{\theta} \mathbf{q} \cdot \text{grad } \theta \equiv \mathbf{j}_c \cdot \mathbf{e}_c + \mathbf{j}_s \cdot \mathbf{e}_s - \frac{1}{\theta} \mathbf{q} \cdot \text{grad } \theta \geq 0. \tag{32}$$

The equivalence of the two forms of dissipation inequality can be proved using the relationships (29)–(31). The choice of spin-dependent variables and the corresponding second law statement presented in this work are consistent with the spintronic formulations in [5, 7, 34, 40]. The thermodynamic driving force vector $\mathcal{F}$, consisting of the complete set of independent variables associated with thermal, electromagnetic, and spin transport processes, is defined as

$$\mathcal{F} = \text{grad } \mathbf{u} = \left\{ \text{grad } \mu_c, \ \text{grad } \mu_s, \ \frac{\text{grad } \theta}{\theta} \right\}. \tag{33}$$

Additional independent variables like the external magnetic field $\mathbf{h}$ or the spin polarization vector $\hat{\sigma}$ maybe required for complete characterization, depending on the physical process. These are usually accommodated within the process constants called *the kinetic coefficients*. The corresponding dependent variables, i.e., *the flow vector* $\mathcal{J}$ is given by

$$\mathcal{J} = \left\{ \mathbf{j}_c, \ \mathbf{j}_s, \ \mathbf{q} \right\}. \tag{34}$$

### 3.2.5. Linear constitutive model

The constitutive equations describing the transport phenomena can now be posited in the general form

$$\mathcal{J}_i = \sum_{j=0}^{n} L_{ij}(\mathbf{h}, \hat{\sigma}) \, \mathcal{F}_j \ + \ \frac{1}{2!} \sum_{j=0}^{n} \sum_{k=0}^{n} L_{ijk}(\mathbf{h}, \hat{\sigma}) \, \mathcal{F}_j \, \mathcal{F}_k \ + \ .... \tag{35}$$

where $\mathcal{J}_i$ and $\mathcal{F}_i$ represent the components of the thermodynamic flow and the thermodynamic force vectors described by (34) and (33), respectively. Also, $\hat{\sigma}$ represents the spin-polarization unit vector and $L_{ij}$, $L_{ijk}$ denote the kinetic coefficients that correlate the fluxes and the driving forces. These coefficients depend on the material property as well as other external factors like applied magnetic field or spin-polarization.

Irreversible transport processes with no memory are known as *Markovian processes* and can be described using only the leading order terms in (35), i.e.,

$$\mathcal{J}_i = \sum_{j=0}^{n} L_{ij}(\mathbf{h}, \hat{\sigma}) \, \mathcal{F}_j. \tag{36}$$

The constitutive equation (36) will now be specialized to linear spin, charge and current transport processes. Since the purely thermoelectric processes occur in the absence of external magnetic fields and spin-polarization, the corresponding kinetic coefficients are assumed to be material specific constants. On the other hand, the kinetic coefficients corresponding to the thermomagnetic and spintronic processes are dependent on external factors like applied magnetic field or spin-polarization. Equation (36) can be considerably simplified for these processes by noting that only the components of magnetic field and spin-polarization vectors orthogonal to both the flow and force quantities contribute to the transport phenomenon, i.e., they only appear as cross-product terms in the constitutive model. The constitutive equations are thus specialized to the form

$$\mathcal{J}_i = \underbrace{L_{ij}^a \, \mathcal{F}_j}_{\text{thermoelectric}} \ + \ \underbrace{L_{il}^b \, \epsilon_{lkj} \, h_k \, \mathcal{F}_j}_{\text{thermomagnetic/galvanomagnetic}} \ + \ \underbrace{L_{il}^c \, \epsilon_{lkj} \, \hat{\sigma}_k \, \mathcal{F}_j}_{\text{spin-induced}} \tag{37}$$

wherein, the constants $L_{ij}^a$ describe the thermoelectric effects. Also, $L_{ij}^b$ represent the process constants corresponding to galvanomagnetic and thermomagnetic effects and $L_{ij}^c$ describe the material constants for the spin-induced processes. Substituting (33)–(34) into (37) and specializing to the transport phenomena highlighted in Figure 2, the constitutive equations are reduced to

$$\mathbf{j}_c = \boldsymbol{\sigma}^c \cdot \mathbf{e}_c + \mathbf{R}^c \cdot \mathbf{h} \times \mathbf{e}_c + \mathbf{R}^s \cdot \hat{\boldsymbol{\sigma}} \times \mathbf{e}_s + \mathbf{S}^c \cdot \frac{\operatorname{grad}\theta}{\theta} + \mathbf{N}^c \cdot \mathbf{h} \times \frac{\operatorname{grad}\theta}{\theta}, \quad (38)$$

$$\mathbf{j}_s = \mathbf{R}^{s'} \cdot \hat{\boldsymbol{\sigma}} \times \mathbf{e}_c + \boldsymbol{\sigma}^s \cdot \mathbf{e}_s + \mathbf{S}^s \cdot \frac{\operatorname{grad}\theta}{\theta} + \mathbf{N}^s \cdot \hat{\boldsymbol{\sigma}} \times \frac{\operatorname{grad}\theta}{\theta}, \quad (39)$$

$$\mathbf{q} = \mathbf{S}^{c'} \cdot \mathbf{e}_c + \mathbf{N}^{c'} \cdot \mathbf{h} \times \mathbf{e}_c + \mathbf{S}^{s'} \cdot \mathbf{e}_s + \mathbf{N}^{s'} \cdot \hat{\boldsymbol{\sigma}} \times \mathbf{e}_s + \mathbf{K} \cdot \frac{\operatorname{grad}\theta}{\theta}$$

$$+ \mathbf{R}^t \cdot \mathbf{h} \times \frac{\operatorname{grad}\theta}{\theta}, \quad (40)$$

wherein all the process constants are described in Table 2. Specializing further to isotropic material, the constitutive equations (38)–(40) can be presented in the matrix form

$$
\begin{bmatrix} j_{c_x} \\ j_{c_y} \\ j_{c_z} \\ j_{s_x} \\ j_{s_y} \\ j_{s_z} \\ q_x \\ q_y \\ q_z \end{bmatrix} =
\begin{bmatrix}
\sigma_{11}^c & -R_{12}^c h_z & R_{12}^c h_y & 0 & -R_{12}^s \hat{\sigma}_z & R_{12}^s \hat{\sigma}_y & S_{11}^c & -N_{12}^c h_z & N_{12}^c h_y \\
R_{21}^c h_z & \sigma_{11}^c & -R_{21}^c h_x & R_{21}^s \hat{\sigma}_z & 0 & -R_{21}^s \hat{\sigma}_x & N_{21}^c h_z & S_{11}^c & -N_{21}^c h_x \\
-R_{31}^c h_y & R_{31}^c h_x & \sigma_{11}^c & -R_{31}^s \hat{\sigma}_y & R_{31}^s \hat{\sigma}_x & 0 & -N_{31}^c h_y & N_{31}^c h_x & S_{11}^c \\
0 & -R_{12}^{s'} \hat{\sigma}_z & R_{12}^{s'} \hat{\sigma}_y & \sigma_{11}^s & 0 & 0 & S_{11}^s & -N_{12}^s \hat{\sigma}_z & N_{12}^s \hat{\sigma}_y \\
R_{21}^{s'} \hat{\sigma}_z & 0 & -R_{21}^{s'} \hat{\sigma}_x & 0 & \sigma_{11}^s & 0 & N_{21}^s \hat{\sigma}_z & S_{11}^s & -N_{21}^s \hat{\sigma}_x \\
-R_{31}^{s'} \hat{\sigma}_y & R_{31}^{s'} \hat{\sigma}_x & 0 & 0 & 0 & \sigma_{11}^s & -N_{31}^s \hat{\sigma}_y & N_{31}^s \hat{\sigma}_x & S_{11}^s \\
S_{11}^{c'} & -N_{12}^{c'} h_z & N_{12}^{c'} h_y & S_{11}^{s'} & -N_{12}^{s'} \hat{\sigma}_z & N_{12}^{s'} \hat{\sigma}_y & K_{11} & -R_{12}^t h_z & R_{12}^t h_y \\
N_{21}^{c'} h_z & S_{11}^{c'} & -N_{21}^{c'} h_x & N_{21}^{s'} \hat{\sigma}_z & S_{11}^{s'} & -N_{21}^{s'} \hat{\sigma}_x & R_{21}^t h_z & K_{11} & -R_{21}^t h_x \\
-N_{31}^{c'} h_y & N_{31}^{c'} h_x & S_{11}^{c'} & -N_{31}^{s'} \hat{\sigma}_y & N_{31}^{s'} \hat{\sigma}_x & S_{11}^{s'} & -R_{31}^t h_y & R_{31}^t h_x & K_{11}
\end{bmatrix}
\begin{bmatrix} e_{c_x} \\ e_{c_y} \\ e_{c_z} \\ e_{s_x} \\ e_{s_y} \\ e_{s_z} \\ (\operatorname{grad}\theta)_x/\theta \\ (\operatorname{grad}\theta)_y/\theta \\ (\operatorname{grad}\theta)_z/\theta \end{bmatrix}.
$$

$$(41)$$

**Table 2.** Process constants for *thermoelectric*, *thermomagnetic* and *spin-induced* effects.

| Thermoelectric | Thermomagnetic | Spin-induced |
|---|---|---|
| $\boldsymbol{\sigma}_{ij}^{c}$ – Electrical Conductivity | $\mathbf{R}_{ij}^{c}$ – Hall Effect | $\mathbf{R}_{ij}^{s}$ – Spin-Hall Effect |
| $\mathbf{K}_{ij}$ – Thermal Conductivity | $\mathbf{R}_{ij}^{c'}$ – Inverse Hall Effect | $\mathbf{R}_{ij}^{s'}$ – Inverse Spin-Hall Effect |
| $\mathbf{S}_{ij}^{c}$ – Seebeck Effect | $\mathbf{N}_{ij}^{c}$ – Nernst Effect | $\boldsymbol{\sigma}_{ij}^{s}$ – Spin Conductivity |
| $\mathbf{S}_{ij}^{c'}$ – Peltier Effect | $\mathbf{N}_{ij}^{c'}$ – Inverse Nernst Effect | $\mathbf{N}_{ij}^{s}$ – Spin-Induced Nernst |
| | $\mathbf{R}_{ij}^{t}$ – Righi–Leduc Effect | $\mathbf{N}_{ij}^{s'}$ – Inverse Spin Nernst |
| | | $\mathbf{S}_{ij}^{s}$ – Spin-Induced Seebeck |
| | | $\mathbf{S}_{ij}^{s'}$ – Spin-Induced Peltier |

# 4. Results and discussion

## 4.1. Restrictions imposed by the Clausius–Duhem inequality and Onsager equations

In what follows, we derive the restrictions imposed by the dissipation inequality (32) and *Onsager Reciprocal Relations* on the system of equations (41).

### *4.1.1. Onsager's reciprocal relations*

Onsager reciprocal relations express the equality of certain ratios between the thermodynamic flows and forces in a linear transport process [8]. Specifically, Onsager relations state that the kinetic coefficients corresponding to these processes can be related as

$$L_{ij} = L_{ji} \quad \text{for } \mathbf{h} = \mathbf{0}, \tag{42}$$

$$L_{ij}(\mathbf{h}) = L_{ji}(-\mathbf{h}) \quad \text{for } i \neq j \text{ and } \mathbf{h} \neq \mathbf{0}. \tag{43}$$

Crystallographic symmetry of the material comes into play when deducing the inverse relations using Onsager equations [2, 3]. Thus, in order to simplify the presentation, isotropic crystal symmetry is assumed here. We now apply these relationships to the process constants in (41).

- The **Hall coefficients** $R_{ij}^c$ are related to each other using (43), i.e.,

$$R_{12}^c h_z = -R_{21}^c(-h_z) \quad \Rightarrow \quad R_{12}^c = R_{21}^c \tag{44}$$

similarly $R_{12}^c = R_{31}^c$. \hfill (45)

  Thus, $R_{12}^c = R_{21}^c = R_{31}^c = R^c$.

- The **Thermal Hall** (or *Righi–Leduc*) coefficients $R_{ij}^t$ can be related in a similar manner, i.e.,

$$R_{12}^t = R_{21}^t = R_{31}^t = R^t. \tag{46}$$

- Onsager relations can also be used to relate inverse processes. For instance, **Seebeck** $S_{ij}^c$ and **Peltier** $S_{ij}^{c'}$ coefficients can be related using (42) as

$$S_{11}^c = S_{11}^{c'} \quad \Rightarrow \quad S^c = S^{c'}. \tag{47}$$

Similarly, the **Nernst** coefficient $N_{ij}^c$ is compared to **Inverse Nernst** coefficient $N_{ij}^{c'}$ as

$$N_{12}^c = N_{21}^c = N_{31}^c = N_{12}^{c'} = N_{21}^{c'} = N_{31}^{c'} \quad \Rightarrow \quad N^c = N^{c'}. \tag{48}$$

- For spin transport processes, we first compare Spin Hall and Inverse Spin Hall coefficients for an isotropic material, using (42), i.e., $R_{ij}^s$ and $R_{ij}^{s'}$ are related as

$$-R_{12}^s \hat{\sigma}_z = R_{21}^{s'} \hat{\sigma}_z \quad \Rightarrow \quad R_{12}^s = -R_{21}^{s'}. \tag{49}$$

Applying similar arguments to all components, we obtain

$$R_{12}^s = R_{21}^s = R_{31}^s = -R_{12}^{s'} = -R_{21}^{s'} = -R_{31}^{s'} \quad \Rightarrow \quad R^s = -R^{s'}. \tag{50}$$

- Finally, comparing the **Spin-induced Seebeck** $S_{ij}^s$ and **Spin Nernst** $N_{ij}^s$ effects to their respective inverses $S_{ij}^{s'}$ and $N_{ij}^{s'}$, we obtain

$$N_{12}^s = N_{21}^s = N_{31}^s = -N_{12}^{s'} = -N_{21}^{s'} = -N_{31}^{s'} \quad \Rightarrow \quad N^s = -N^{s'} \tag{51}$$

$$\text{Finally,} \quad S_{11}^s = S_{11}^{s'} \quad \Rightarrow \quad S^s = S^{s'}. \tag{52}$$

Thus, the Onsager relationships reduce all the kinetic coefficients for an isotropic material to scalar quantities. The transport equations (38)–(40) are thus reduced to

$$\mathbf{j}_c = \sigma^c \, \mathbf{e}_c + R^c \, \mathbf{h} \times \mathbf{e}_c + R^s \, \hat{\sigma} \times \mathbf{e}_s + S^c \, \frac{\operatorname{grad}\theta}{\theta} + N^c \, \mathbf{h} \times \frac{\operatorname{grad}\theta}{\theta}, \tag{53}$$

$$\mathbf{j}_s = -R^s \, \hat{\sigma} \times \mathbf{e}_c + \sigma^s \, \mathbf{e}_s + S^s \, \frac{\operatorname{grad}\theta}{\theta} + N^s \, \hat{\sigma} \times \frac{\operatorname{grad}\theta}{\theta}, \tag{54}$$

$$\mathbf{q} = S^c \, \mathbf{e}_c + N^c \, \mathbf{h} \times \mathbf{e}_c + S^s \, \mathbf{e}_s - N^s \, \hat{\sigma} \times \mathbf{e}_s + K \, \frac{\operatorname{grad}\theta}{\theta} + R^t \, \mathbf{h} \times \frac{\operatorname{grad}\theta}{\theta}. \tag{55}$$

## 4.2. Restrictions imposed by the second law of thermodynamics

The dissipation inequality (32) is now imposed on the reduced constitutive equations (53)–(55). The dissipation inequality is rewritten as

$$\Gamma(\mathcal{F}) \equiv \mathbf{j}_c(\mathcal{F}) \cdot \mathbf{e}_c + \mathbf{j}_s(\mathcal{F}) \cdot \mathbf{e}_s - \mathbf{q}(\mathcal{F}) \cdot \frac{\operatorname{grad}\theta}{\theta} \geq 0 \tag{56}$$

wherein equality occurs only at equilibrium. Thus, at equilibrium the function $\Gamma(\mathcal{F})$ is minimized with respect to the independent variables $e_{c_i}$, $e_{s_i}$ and $\theta_{,i}/\theta$, i.e.,

$$\frac{\partial \Gamma}{\partial \mathbf{e}_c}\Big|_E = \frac{\partial \Gamma}{\partial \mathbf{e}_s}\Big|_E = \frac{\partial \Gamma}{\partial(\operatorname{grad}\theta/\theta)}\Big|_E = \mathbf{0}, \tag{57}$$

where $()|_E$ denotes the value of the enclosed quantity at equilibrium. Substituting (53)–(55) into the equilibrium conditions (57) and solving the resulting system of equations, we obtain

$$\mathbf{e}_c|_E = \mathbf{e}_s|_E = (\operatorname{grad}\theta)|_E = \mathbf{0} \quad \Leftrightarrow \quad \mathbf{j}_c|_E = \mathbf{j}_s|_E = \mathbf{q}|_E = \mathbf{0} \tag{58}$$

at equilibrium. Rewriting dissipation inequality (56) using constitutive equations (53)–(55) and rearranging the terms

$$\Gamma(\mathcal{F}) \equiv \mathbf{j}_c(\mathcal{F}) \cdot \mathbf{e}_c + \mathbf{j}_s(\mathcal{F}) \cdot \mathbf{e}_s - \mathbf{q}(\mathcal{F}) \cdot \frac{1}{\theta} \operatorname{grad}\theta \geq 0 \tag{59}$$

$$\Rightarrow \left( \sigma^c \, \mathbf{e}_c + R^c \, \mathbf{h} \times \mathbf{e}_c + R^s \, \hat{\sigma} \times \mathbf{e}_s + S^c \, \frac{\operatorname{grad}\theta}{\theta} + N^c \, \mathbf{h} \times \frac{\operatorname{grad}\theta}{\theta} \right) \cdot \mathbf{e}_c$$

$$+ \left( -R^s \, \hat{\sigma} \times \mathbf{e}_c + \sigma^s \, \mathbf{e}_s + S^s \, \frac{\operatorname{grad}\theta}{\theta} + N^s \, \hat{\sigma} \times \frac{\operatorname{grad}\theta}{\theta} \right) \cdot \mathbf{e}_s$$

$$- \left( S^c \, \mathbf{e}_c + N^c \, \mathbf{h} \times \mathbf{e}_c + S_s \, \mathbf{e}_s - N_s \, \hat{\sigma} \times \mathbf{e}_s \right.$$

$$+ K\,\frac{\operatorname{grad}\theta}{\theta} + R^t\,\mathbf{h}\times\frac{\operatorname{grad}\theta}{\theta}\Bigg)\cdot\frac{\operatorname{grad}\theta}{\theta}\ \geq\ 0$$

$$\Rightarrow\quad \sigma^c\,|\mathbf{e}_c|^2 + \sigma^s\,|\mathbf{e}_s|^2 - K\,\left|\frac{\operatorname{grad}\theta}{\theta}\right|^2 + 2\,N^c\left(\mathbf{h}\times\frac{\operatorname{grad}\theta}{\theta}\right)\cdot\mathbf{e}_c$$

$$+\,2\,R^s\left(\hat{\boldsymbol{\sigma}}\times\mathbf{e}_s\right)\cdot\mathbf{e}_c\ \geq\ 0. \tag{60}$$

The inequality (60) must be valid for any arbitrary values of $\mathbf{e}_c$, $\mathbf{e}_s$ and $(\operatorname{grad}\theta)/\theta$. This is utilized to derive the restrictions on kinetic coefficients using the techniques presented in Section 3.1.2. The inequality (32) is investigated for the following special cases:

- Case I: For $\mathbf{e}_c \neq \mathbf{0}$, $\mathbf{e}_s = \mathbf{0}$ and $\operatorname{grad}\theta = \mathbf{0}$ the inequality reduces to

$$\sigma^c\,|\mathbf{e}_c|^2\ \geq\ 0 \qquad \Rightarrow \qquad \sigma^c\ \geq\ 0. \tag{61}$$

Restrictions on spin conductivity and thermal conductivity are derived using similar arguments, i.e.,

$$\sigma^s\ \geq\ 0, \qquad K\ \leq\ 0. \tag{62}$$

Since thermal conductivity $K$ is always less than or equal to zero, we define

$$\kappa = -K,\ \text{such that}\ \kappa\ \geq\ 0. \tag{63}$$

- Case II: When any two or more of the three independent variables are non-zero, the inequality (60) can be rewritten as

$$\Gamma(\mathcal{F})\ \equiv\ \frac{1}{\sigma^s}\left(\sigma^s\,\mathbf{e}_s - R^s\left(\hat{\boldsymbol{\sigma}}\times\mathbf{e}_c\right)\right)^2 + \frac{1}{\kappa}\left(\kappa\,\frac{\operatorname{grad}\theta}{\theta} - N^c\left(\mathbf{h}\times\mathbf{e}_c\right)\right)^2$$

$$+\,\left(\sigma^c - \frac{(R^s)^2}{\sigma_s}\,|\hat{\boldsymbol{\sigma}}|^2\sin^2\phi_1 - \frac{(N^c)^2}{\kappa}\,|\mathbf{h}|^2\sin^2\phi_2\right)|\mathbf{e}_c|^2\ \geq\ 0, \tag{64}$$

where $\phi_1$ is the angle between the spin-polarization vector $\hat{\boldsymbol{\sigma}}$ and the charge electric field $\mathbf{e}_c$ and $\phi_2$ is the angle between the magnetic field $\mathbf{h}$ and $\mathbf{e}_c$. Also, the magnitude of spin polarization vector $|\hat{\boldsymbol{\sigma}}| = 1$ and $\kappa$ is defined by (63). In order for the inequality (64) to hold for any values of $\mathbf{e}_c$, $\mathbf{e}_s$ and $\operatorname{grad}\theta/\theta$ the coefficients of the square terms must be always positive. This implies

$$\sigma^c - \frac{(R^s)^2}{\sigma^s}\sin^2\phi_1 - \frac{(N^c)^2}{\kappa}\,|\mathbf{h}|^2\sin^2\phi_2\ \geq\ 0$$

$$\Rightarrow \sigma^c\ \geq\ \frac{(R^s)^2}{\sigma^s}\sin^2\phi_1 + \frac{(N^c)^2}{\kappa}\,|\mathbf{h}|^2\sin^2\phi_2 \qquad \forall\phi_1,\phi_2$$

$$\Rightarrow \sigma^c\ \geq\ \frac{(R^s)^2}{\sigma^s} + \frac{(N^c)^2}{\kappa}\,|\mathbf{h}|^2, \tag{65}$$

along with the conditions derived in (61)–(62).

## 4.3. A note on the governing equations for transport phenomena

In order to utilize the thermodynamic framework for transport device modeling, the constitutive equations need to be supplemented with the appropriate governing equations and boundary conditions. For instance, the complete model for heat transport (in the absence of charge or spin transport) includes the constitutive equation for heat conduction, the conservation of energy (1d) and the appropriate material boundary conditions (e.g., insulation). Similarly, for charge transport *conservation of charge* is invoked, which is a result of the combination of the Gauss's law for electricity (1h) and the Ampère–Maxwell law (1i):

$$\dot{\sigma}_c + \operatorname{div} \mathbf{j}_c = 0,$$

which reduces to

$$\operatorname{div} \mathbf{j}_c = 0 \qquad (66)$$

in the absence of time varying charge density.

**Spin transport**: Spin transport differs from charge transport in that spin is a nonconserved quantity in solids due to the spin-flip mechanism of decay for a spin-polarized electron population. Evolution of spin-voltage is instead described through the phenomenological Valet–Fert equation [34]

$$\nabla^2 \mu_s = \frac{\mu_s}{\lambda_F{}^2} \qquad (67)$$

where $\lambda_F$ is the spin-flip diffusion length and $\mu_s$ is the spin voltage.

**Boundary conditions**: At the free or vacuum interface the spin current vanishes, i.e., $\mathbf{j}_s^{(V)} = \mathbf{0}$. Across the nonmagnetic (NM) and ferromagnetic (FM) boundary, the spin current is described as

$$\mathbf{j}_s^{(F)} = G_r \, \hat{\mathbf{m}} \times \frac{d\,\hat{\mathbf{m}}}{d\,t} \qquad (68)$$

where $\hat{\mathbf{m}}$ is the magnetic moment vector and $G_r$ is the real part of the spin-mixing conductance at the NM|FM interface [1].

## 5. Conclusion

A unified thermodynamic framework was developed for the characterization of functional materials exhibiting *thermo-electro-magneto-mechanical* (TEMM) behavior. Particularly, this overarching framework combines electrodynamics of continua, classical equilibrium and non-equilibrium thermodynamics principles to enable the characterization of a broad range of linear reversible and irreversible TEMM processes highlighted in Figures 1–2.

In the first part of the paper, starting from the state equations presented in [33], a constitutive modeling framework was developed for a fully coupled *reversible* (or quasistatic) TEMM system. Stability conditions were further imposed on the resulting internal energy function. The utility of this framework was demonstrated by specializing the TEMM material to a multiferroic with hexagonal crystal symmetry and subsequently deducing the bounds on the material constants. In the second part of the paper, principles of irreversible thermodynamics were used to develop constitutive models for linear charge, heat and *spin* transport phenomena wherein the dissipation inequality was modified to incorporate the spin-polarized physical quantities. As a result of this modification, in addition to the standard thermoelectric, thermomagnetic and galvanomagnetic transport phenomena, the characterization of spintronic and spin caloritronic effects emerged as a part of this formalism. Onsager's reciprocal relations and second law of thermodynamics were invoked to deduce bounds on the kinetic coefficients.

Applications of this framework are envisioned in design and characterization of functional materials. Also, the restrictions derived in this work, like (17)–(22) and (61)–(65), can be imposed as design constraints while optimizing the material properties.

## Declarations

### Author contribution statement

Sushma Santapuri: Conceived and designed the analysis; Analyzed and interpreted the data; Contributed analysis tools or data; Wrote the paper.

### Funding statement

This work was supported by Institute for Functional Nanomaterials, University of Puerto Rico and NSF grant (EPS-1002410).

### Competing interest statement

The authors declare no conflict of interest.

### Additional information

No additional information is available for this paper.

## Acknowledgements

The author would like to thank Professor Joseph Heremans, Professor Stephen Bechtel, Dr. Robert Lowe and Professor Ernesto Ulloa for valuable discussions and input.